\documentclass[conference]{IEEEtran}
\IEEEoverridecommandlockouts
\usepackage{cite}
\usepackage{amsmath,amssymb,amsfonts}
\usepackage{algorithmic}
\usepackage{graphicx}
\usepackage{textcomp}
\usepackage{xcolor}
\usepackage{booktabs}
\usepackage{algorithm}
\usepackage{hyperref}
\hypersetup{hypertex=true,
	colorlinks=true,
	linkcolor=black,
	anchorcolor=black,
	citecolor=black}
\def\BibTeX{{\rm B\kern-.05em{\sc i\kern-.025em b}\kern-.08em
    T\kern-.1667em\lower.7ex\hbox{E}\kern-.125emX}}

\begin{document}

\title{Reinforcement Solver for H-infinity Filter with Bounded Noise\\
\thanks{This work is supported by International Science \& Technology Cooperation Program of China under 2019YFE0100200, and Beijing Natural Science Foundation with JQ18010. All correspondence should be sent to S. Li with email: lisb04@gmail.com.}
}

\author{\IEEEauthorblockN{Jie Li, Shengbo Eben Li$^*$, Kaiming Tang, Yao Lv, Wenhan Cao}
\IEEEauthorblockA{\textit{School of Vehicle and Mobility, Tsinghua University, Beijing, 100084, China} \\
Email: \{jie-li18, tkm18, y-lv19, cwh19\}@mails.tsinghua.edu.cn, lisb04@gmail.com
}}

\maketitle

\begin{abstract}
H-infinity filter has been widely applied in engineering field, but copping with bounded noise is still an open problem and difficult to solve. This paper considers the H-infinity filtering problem for linear system with bounded process and measurement noise. The problem is first formulated as a zero-sum game where the dynamic of estimation error is non-affine with respect to filter gain and measurement noise. A nonquadratic Hamilton-Jacobi-Isaacs (HJI) equation is then derived by employing a nonquadratic cost to characterize bounded noise, which is extremely difficult to solve due to its non-affine and nonlinear properties. Next, a reinforcement learning algorithm based on gradient descent method which can handle nonlinearity is proposed to update the gain of reinforcement filter, where measurement noise is fixed to tackle non-affine property and increase the convexity of Hamiltonian. Two examples demonstrate the convergence and effectiveness of the proposed algorithm.
\end{abstract}


\section{Introduction}
\label{sec::introduction}

It is well known that filtering techniques are applicable to almost all areas of engineering and science, and there is extensive research on filtering in the field of signal processing and control since the successful application of Kalman filter in space projects in the 1960s \cite{c1}\cite{c2}. H-infinity filter aims to bound the maximum of the energy gain which relates the noise with the estimation error of a linear combination of states, i.e., the $H_\infty$ norm of the system \cite{c3}. In the $H_\infty$ setting, the noise is assumed to be active, and it may seek to reduce the accuracy of estimation; while the filter is committed to reducing estimation error. Through this game mechanism, $H_\infty$ filter performs more robustly when there exists uncertainty in the system compared with Kalman filter, and it is not required to make any statistic assumptions about process noise and measurement noise.

Basic researches for $H_\infty$ filter have been conducted including continuous- and discrete-time cases, time-varying and time-invariant systems, which are summarized in \cite{c3}\cite{c4}. A Riccati equation-based approach to solving $H_\infty$ filtering in the common case where process and measurement noise is not energy bounded is provided in \cite{c5}. Robust $H_\infty$ estimation guarantees that the $H_\infty$ norm of the system is below a given attenuation level for all admissible uncertainties \cite{c6}. Efficient linear matrix inequality (LMI) approaches are provided to solve robust $H_\infty$ filters with uncertainties described by integral quadratic constraints (IQCs), which are suitable to characterize other signal processing applications such as uncertain parameters and time delays\cite{c7}. However, few studies have address $H_\infty$ filtering problems with bounded noise, which is a common object in engineering practice.

For nonlinear systems, a fuzzy $H_\infty$ filter relieving the ill-conditioning caused by the interaction of slow and fast dynamic modes is developed in \cite{c8}, where the perturbed dynamic is described by a multimodel approach. Linear approximation methods can be employed to design an extended robust $H_\infty$ filter \cite{c9} for general nonlinear systems with noise described by an IQC. By embedding the unscented transform technique into the extended $H_\infty$ filter structure, the unscented $H_\infty$ filtering can be carried out using the statistical linear error propagation approach, which achieves not only higher accuracy, but also robustness against model uncertainty compared with the extended $H_\infty$ filter and the unscented Kalman filter \cite{c10}.

Studies mentioned above employ LMI-based or algebraic Riccati equation (ARE)-based approaches and linear approximation methods to cope with nonlinear dynamics. Directly dealing with nonlinear properties and bounded noise requires further research. Reinforcement learning is an class of effective methods to solve the optimal control problem of nonlinear systems \cite{ce1}\cite{ce2}. As one of the model-based methods, policy iteration is an iterative technique to solve HJI equation which is a nonlinear version of ARE for solving $H_\infty$ filter and $H_\infty$ control \cite{c11}\cite{c12}. The policy iteration framework involves two steps. Policy evaluation step evaluates the value function of given policies, and policy improvement step updates policies by minimizing or maximizing Hamiltonian. And a nonquadratic functional enables policy iteration to confront control saturation in $H_\infty$ state-feedback and tracking control \cite{c11}\cite{c13}. Nevertheless, the improved policies can not be explicitly calculated for non-affine systems, because the minimize and maximize operations in HJI equation do not separate out \cite{c14}, which brings difficulties to algorithm design and practical calculation.

This study presents a reinforcement $H_\infty$ filter subject to bounded process and measurement noise. From the perspective of control theory, the $H_\infty$ filtering problem for a linear system can be formulated as a zero-sum game with non-affine dynamics, but corresponding HJI equation can not be solved by existing policy iteration methods. Inspired by our previous work \cite{c15}, a ternary policy iteration (TPI) algorithm successfully solves it numerically, where a fixed measurement noise technique is used to increase the convexity of filter gain, which also provides new ideas for handling nonlinear systems.

This paper is organized as follows. Section \ref{sec::problem description} describes the $H_\infty$ filtering problem considering bounded noise. The problem is formulated as a zero-sum game, and corresponding nonquadratic HJI equation is derived in Section \ref{sec::nonquadratic HJI equation for H_infty filtering with bounded noise}. Section \ref{sec::degenerate into traditional methods regardless of the bounded property of noise} shows that the solution of nonquadratic HJI equation can degenerate into that of traditional $H_\infty$ filter. Section \ref{sec::ternary policy iteration algorithm for H_infty filter} proposes the TPI algorithm. Two examples are presented in Section \ref{sec::illustrative examples}. And Section \ref{sec::conclusions} gives a conclusion in the end.

\textbf{Notation}~~The notation employed throughout this study is relatively standard. $\|x\|_2 = \sqrt{x^T x}$ denotes the 2-norm of the vector $x$, and $\|x\|_Q^2 = x^T Q x$. $\|A\|_F = \sqrt{\sum_{i=1}^{m}\sum_{j=1}^{n}a_{ij}^2}$ is the Frobenius norm of the matrix $A \in \mathbb{R}^{m \times n}$ where $\mathbb{R}$ denotes the set of real numbers. ${\rm diag}(\cdot)$ refers a diagonal matrix, and $\tanh^{-T}(s) = \left[\tanh^{-1}(s_1), \cdots, \tanh^{-1}(s_n)\right] \in \mathbb{R}^{1 \times n}$ refers a row vector where $s \in \mathbb{R}^n$. $\mathbb{E}_{x\in\mathcal{D}} \left[f(x)\right]$ means expectation of $f(x)$ defined in the set $\mathcal{D}$.

\section{Problem Description}
\label{sec::problem description}

In this section, we will introduce the designing objective of $H_\infty$ filter for linear system with bounded noise in infinite horizon where the initial time $t_0 = 0$. A nonquadratic function is employed to characterize the bounded attribute of noise.

Consider the following time-invariant linear system:
\begin{equation}
	\label{eq::linear system}
	\left\{
	\begin{array}{c}
		\dot{x} = Ax + Bu + w \\
		y = Cx + v \\
		z = Lx
	\end{array}\right.
\end{equation}
where $x \in \mathbb{R}^n$ is the state, $u \in \mathbb{R}^m$ is the control variable, $w \in \mathbb{R}^n$ is bounded process noise and $|w| \leq \bar{w} \in \mathbb{R}^n$, $y \in \mathbb{R}^r$ is the measurement, $v \in \mathbb{R}^r$ is bounded measurement noise and $|v| \leq \bar{v} \in \mathbb{R}^r$, $z \in \mathbb{R}^s$ is the output to be estimated.

The aim of $H_\infty$ filter in infinite horizon is to find an estimate $\hat{z}$ of $z$ to suppress the $L_2$ gain with respect to noise and estimation error below a self-defined attenuation level $\gamma$ \cite{c5}
\begin{equation}
	\label{eq::objective of H_infty filter}
	\frac{\int_0^\infty \|z - \hat{z}\|_S^2 dt}{\int_0^\infty \left(\mathcal{F}(w) + \mathcal{F}(v)\right) dt} \leq \gamma^2, \forall w,v
\end{equation}
where noise is depicted with nonquadratic terms \cite{c11}\cite{c13}
\begin{equation}
	\label{eq::nonquadratic terms}
	\begin{array}{c}
		\displaystyle{\mathcal{F}(w) = 2\int_0^w \tanh^{-T}\left(\bar{W}^{-1}s\right)\bar{W}Qds} \\
		\displaystyle{\mathcal{F}(v) = 2\int_0^v \tanh^{-T}\left(\bar{V}^{-1}s\right)\bar{V}Rds}
	\end{array}
\end{equation}
in which $\bar{W} = {\rm diag}(\bar{w}) \in \mathbb{R}^{n \times n}$ and $\bar{V} = {\rm diag}(\bar{v}) \in \mathbb{R}^{r \times r}$, estimation error weighting matrix $S \in \mathbb{R}^{s \times s}$, noise weighting matrices $Q \in \mathbb{R}^{n \times n}$ and $R \in \mathbb{R}^{r \times r}$ are all symmetric positive definite, and $\mathcal{F}(\cdot)$ is a bounded even function.

\textbf{Remark 1}~~The reason why it is called $H_\infty$ filter is that when noise is depicted with 2-norm and $S$ is an identity matrix with appropriate dimension, the supremum of the above $L_2$ gain is actually the $H_\infty$ norm of the transform function from noise $e = \left[w^T \quad v^T\right]^T$ to estimation error $\tilde{z} = z - \hat{z}$, i.e.,
\begin{equation}
	\nonumber
	\left\|T_{\tilde{z}e}\right\|_\infty^2 = \displaystyle{\sup_e}\frac{\int_0^\infty \|\tilde{z}\|_2^2 dt}{\int_0^\infty \|e\|_2^2 dt} = \displaystyle{\sup_e}\frac{\int_0^\infty \|z - \hat{z}\|_2^2 dt}{\int_0^\infty \left(\|w\|_2^2 + \|v\|_2^2\right) dt}
\end{equation}

\section{Nonquadratic HJI Equation for H-infinity Filter with Bounded Noise}
\label{sec::nonquadratic HJI equation for H_infty filtering with bounded noise}

In this section, the filtering problem is reconsidered from the perspective of control theory and a zero-sum game is constructed. Then, HJI equation considering bounded noise is derived, whose solution method will be introduced later.

\subsection{Zero-Sum Game of Linear Estimator}

Although in the traditional derivation of $H_\infty$ filter, its specific form is not limited, the form of filter in this section is limited to a linear one, which is the result of $H_\infty$ filter and Kalman filter. And it has the following general expression \cite{c16}
\begin{equation}
	\nonumber
	\dot{\hat{x}} = K_1\hat{x} + Ky + K_3u
\end{equation}
where $\hat{x}$ is an estimate of $x$, $K_1 \in \mathbb{R}^{n \times n}$, $K \in \mathbb{R}^{n \times r}$ and $K_3 \in \mathbb{R}^{n \times m}$ are matrices to be designed.

Define the state estimate error as $\tilde{x} = x - \hat{x}$, assume that the estimation error can be maintained at zero equilibrium point when there is no noise, i.e., when $\tilde{x} = 0$, $w = 0$ and $v = 0$, it has $\dot{\tilde{x}} = 0$. Under this requirement, the linear estimator can be formulated as
\begin{equation}
	\nonumber
	\dot{\hat{x}} = A\hat{x} + Bu + K(y - C\hat{x})
\end{equation}
which has the same expression as that of $H_\infty$ filter and Kalman filter except undetermined filter gain $K$.

Therefore, the dynamic of the state estimate error is
\begin{equation}
	\label{eq::dynamic of the state estimate error}
	\dot{\tilde{x}} = (A - KC)\tilde{x} + w - Kv
\end{equation}
which is considered as the controlled system, where the process noise $w$ and measurement noise $v$ of the original system can be regarded as the noise of the new system dynamics, and the multiplication of $K$ and $v$ leads to a non-affine property.

The designing objective (\ref{eq::objective of H_infty filter}) of $H_\infty$ filter is equivalent to finding an estimate $\hat{z}$ so that
\begin{equation}
	\nonumber
	\displaystyle{\max_{w,v}} \int_0^\infty \|z - \hat{z}\|_S^2 - \gamma^2 \left(\mathcal{F}(w) + \mathcal{F}(v)\right) dt \leq 0
\end{equation}

Substituting $z = Lx$ and $\hat{z} = L\hat{x}$ into the above integral index yields the following value function and performance
\begin{equation}
	\label{eq::value and performance with bounded noise}
	V(\tilde{x}) = J(\tilde{x}, K, v, w) = \int_0^\infty l(\tilde{x}, K, v, w) dt
\end{equation}
with the utility function defined as
\begin{equation}
	\label{eq::nonquadratic utility}
	l(\tilde{x}, K, v, w) = \|\tilde{x}\|_{L^T S L}^2 - \gamma^2\left(\mathcal{F}(w) + \mathcal{F}(v)\right)
\end{equation}

There are many filters that can attain the given attenuation level $\gamma$. A feasible solution is to minimize the estimation error $\tilde{x}$ with respect to the above performance after maximizing noise $v$ and $w$, which is equivalent to minimizing filter gain $K$ because it dominates the estimation error. From the perspective of game theory, it can be formulated as a differential game
\begin{equation}
	\label{eq::zero-sum game with bounded noise}
	\begin{array}{c}
		V^*(\tilde{x}) = \displaystyle{\min_{K} \max_{w,v}} J(\tilde{x}, K, v, w) \\
		s.t. \quad \dot{\tilde{x}} = (A - KC)\tilde{x} + w - Kv
	\end{array}
\end{equation}
where $V^*(\tilde{x})$ is the optimal value function. The filter gain and noise are placed in opposite positions so that one loses what the other gains in its performance. Thus, it is a zero-sum game.

\subsection{Nonquadratic HJI Equation Considering Bounded Noise}

A necessary condition for the existence of the unique solution to the above zero-sum game is Nash condition \cite{c17}
\begin{equation}
	\nonumber
	\displaystyle{\min_{K} \max_{w,v}} J(\tilde{x}, K, v, w) = \displaystyle{\max_{w,v} \min_{K}} J(\tilde{x}, K, v, w)
\end{equation}

Take partial derivative of the initial time $t_0 = 0$ on both sides of value function (\ref{eq::value and performance with bounded noise}), and obtain the Bellman equation about Hamiltonian
\begin{equation}
	\label{eq::Hamiltonian with bounded noise}
	\begin{array}{c}
	\displaystyle{H\bigg(\tilde{x}, K, v, w, \frac{\partial V(\tilde{x})}{\partial \tilde{x}}\bigg) = \|\tilde{x}\|_{L^T S L}^2 - \gamma^2\mathcal{F}(w)} \\
	\displaystyle{- \gamma^2\mathcal{F}(v)
	+ \frac{\partial V(\tilde{x})}{\partial \tilde{x}^T}\left((A - KC)\tilde{x} + w - Kv\right)} = 0
	\end{array}
\end{equation}

Then, a necessary condition for Nash condition is Isaacs’ condition \cite{c17}, which can be regarded as an extension of Pontryagin maximum principle, i.e.
\begin{equation}
	\nonumber
	\min_{K}\max_{w,v}H\!\left(\!\tilde{x},\!K,\!v,\!w,\!\frac{\partial V\!(\!\tilde{x}\!)}{\partial \tilde{x}}\!\right)\!=\!\max_{w,v}\min_{K} H\!\left(\!\tilde{x},\!K,\!v,\!w,\!\frac{\partial V\!(\!\tilde{x}\!)}{\partial \tilde{x}}\!\right)
\end{equation}

Substituting the saddle point into the Bellman equation (\ref{eq::Hamiltonian with bounded noise}) yields the HJI equation which should be a partial differential equation about the optimal value function $V^*(\tilde{x})$
\begin{equation}
	\label{eq::HJI equation}
	\displaystyle{\min_{K} \max_{w, v}} H\left(\tilde{x}, K, v, w, \frac{\partial V^*(\tilde{x})}{\partial \tilde{x}}\right) = 0
\end{equation}

Apply stationary conditions $\partial H/\partial w = 0$, $\partial H/\partial v = 0$ and $\partial H/\partial K = 0$ to the HJI equation (\ref{eq::HJI equation}) to obtain the expressions of the worst-case noise and the optimal filter gain
\begin{equation}
	\label{eq::bounded noise w and v}
	\begin{array}{c}
		\displaystyle{w^* = \bar{W}\tanh\left(\frac{1}{2\gamma^2}(Q\bar{W})^{-1} \frac{\partial V^*(\tilde{x})}{\partial \tilde{x}}\right)} \\
		\displaystyle{v^* = -\bar{V}\tanh\left(\frac{1}{2\gamma^2}(R\bar{V})^{-1} {K^*}^T \frac{\partial V^*(\tilde{x})}{\partial \tilde{x}}\right)} \\
		\displaystyle{\frac{\partial V^*(\tilde{x})}{\partial \tilde{x}^T}}(C\tilde{x} + v^*)^T = 0
	\end{array}
\end{equation}
where $w^*$ and $v^*$ are bounded by boundary values $\bar{w}$ and $\bar{v}$ according to the nature of hyperbolic tangent function $\tanh(\cdot)$.

Note that $v^*$ is not only a function of $V^*(\tilde{x})$ but also $K^*$,
and $K^*$ itself has no specific expression. In order to find the expression of $K^*$ and increase its convexity, fix the noise $v(K) = -\bar{V}\tanh\left(\frac{1}{2\gamma^2}(R\bar{V})^{-1} K^T \frac{\partial V(\tilde{x})}{\partial \tilde{x}}\right)$ in the HJI equation and derive the nonquadratic HJI equation as follows
\begin{equation}
	\label{eq::nonquadratic HJI equation}
	\displaystyle{\min_{K} \max_{w}} H\left(\tilde{x}, K, v(K), w, \frac{\partial V^*(\tilde{x})}{\partial \tilde{x}}\right) = 0
\end{equation}
which is the key to cope with bounded noise and employs nonquadratic terms (\ref{eq::nonquadratic terms}).

In the next section, we will show that it is an extension of the traditional solution method of $H_\infty$ filter, and we will present a policy iteration algorithm to solve the nonquadratic HJI equation in a numerical way in Section \ref{sec::ternary policy iteration algorithm for H_infty filter}.

\section{Degenerate into Traditional Methods Regardless of the Bounded Property of Noise}
\label{sec::degenerate into traditional methods regardless of the bounded property of noise}

In the setting of traditional $H_\infty$ filter, the nonquadratic terms used to depict noise are replaced by 2-norm terms. Therefore, the designing objective is to find an estimate $\hat{z}$ so that
\begin{equation}
	\nonumber
	\displaystyle{\max_{w,v}} \int_0^\infty \|z - \hat{z}\|_S^2 - \gamma^2 \left(\|w\|_{Q^{-1}}^2 + \|v\|_{R^{-1}}^2\right) dt \leq 0
\end{equation}

\textbf{Remark 2}~~Note that $v = y - Cx$, the zero-sum game applied to solve traditional $H_\infty$ filter is defined as
\begin{equation}
	\nonumber
	\begin{array}{c}
	\displaystyle{\min_{\hat{x}} \max_{w,y}} \int_0^\infty \|x - \hat{x}\|_{L^T S L}^2 - \gamma^2\left(\|w\|_{Q^{-1}}^2 + \|v\|_{R^{-1}}^2\right) dt \\
	s.t. \quad \dot{x} = Ax + Bu + w
	\end{array}
\end{equation}
where the optimization order of $\hat{x}$, $y$ and $w$ may affect the solution result of the game problem \cite{c4}. Generally, maximize the process noise $w$ first to obtain the worst-case process noise $w^*$ and corresponding state trajectory $x^*$, which is ``the most difficult'' one to estimate. Then, $w^*$ and $x^*$ are used to perform minimax optimization with respect to state estimation $\hat{x}$ and measurement $y$ to obtain an estimator which is a function of measurement $y$ \cite{c5}.

Similar to the previous section, Isaacs’ condition still holds. The Bellman equation can be written as
\begin{equation}
	\label{eq::Bellman equation}
	\begin{array}{c}
	\displaystyle{H\bigg(\tilde{x}, K, v, w, \frac{\partial V(\tilde{x})}{\partial \tilde{x}}\bigg) = \|\tilde{x}\|_{L^T S L}^2 - \gamma^2\|w\|_{Q^{-1}}^2} \\
	\displaystyle{- \gamma^2\|v\|_{R^{-1}}^2\!+\!\frac{\partial V(\tilde{x})}{\partial \tilde{x}^T}\left((A - KC)\tilde{x} + w - Kv\right)} = 0
	\end{array}
\end{equation}
and the corresponding HJI equation is the same as that in (\ref{eq::HJI equation}).

For the HJI equation, applying stationary conditions $\partial H/\partial w = 0$, $\partial H/\partial v = 0$ and $\partial H/\partial K = 0$ gives
\begin{equation}
	\label{eq::noise w and v}
	\begin{array}{c}
	\displaystyle{w^* = \frac{1}{2\gamma^2} Q \frac{\partial V^*(\tilde{x})}{\partial \tilde{x}}} \\
	\displaystyle{v^* = -\frac{1}{2\gamma^2} R{K^*}^T \frac{\partial V^*(\tilde{x})}{\partial \tilde{x}}} \\
	\displaystyle{\frac{\partial V^*(\tilde{x})}{\partial \tilde{x}^T}}(C\tilde{x} + v^*)^T = 0
	\end{array}
\end{equation}
where $w^*$ and $v^*$ satisfy the saddle point condition because their second-order partial derivatives are negative definite.

Similarly, the above formula only gives a relationship between $v^*$ and $K^*$, and the filter gain $K^*$ is not a local minimum with respect to Hamiltonian because its second-order partial derivative is not positive. 

By substituting $w^*$ and $v^*$ into (\ref{eq::Bellman equation}) and analyzing the structure of the obtained Hamiltonian $H\left(\tilde{x}, K, v^*, w^*, \frac{\partial V(\tilde{x})}{\partial \tilde{x}}\right)$, it can be inferred that $V^*(\tilde{x})$ is a quadratic form of $\tilde{x}$. So the optimal value function is set as
\begin{equation}
	\label{eq::value function}
	V^*(\tilde{x}) = \gamma^2\tilde{x}^TP^{-1}\tilde{x}
\end{equation}
where $P^{-1}$ is positive definite. Then, the worst-case noise $w^*$ and $v^*$ can be rewritten as
\begin{equation}
	\label{eq::w and v}
	\begin{array}{c}
		w^* = QP^{-1}\tilde{x} \\
		v^* = -R{K^*}^T P^{-1}\tilde{x}
	\end{array}
\end{equation}

In order to derive the optimal gain $K^*$ and ensure that its second-order partial derivative is positive, fix the measurement noise $v(K)\!=\!-RK^TP^{-1}\tilde{x}$. Substitute $v(K)$ and value function (\ref{eq::value function}) into (\ref{eq::Bellman equation}). Then, the obtained Hamiltonian $H\left(\tilde{x}, K, v(K), w, P\right)$ is a quadratic function relating to filter gain $K$. Reapply stationary condition $\partial H/\partial K = 0$ and note the positive definiteness of $P^{-1}$, one can get the optimal gain
\begin{equation}
	\label{eq::the optimal gain}
	K^* = PC^TR^{-1}
\end{equation}

Suppose there is a small disturbance $\alpha\Delta K$ at $K^*$, the corresponding Hamiltonian can be derived as
\begin{equation}
	\nonumber
	\begin{array}{c}
	H\left(\tilde{x}, K^* + \alpha\Delta K, v(K^* + \alpha\Delta K), w, P\right) \\
	= H\left(\tilde{x}, K^*, v(K^*), w, P\right) + \alpha^2\gamma^2\left\|\Delta K^TP^{-1}\tilde{x}\right\|_R^2
	\end{array}
\end{equation}
which means $K^*$ is a local minimum and satisfies the saddle point condition.

Finally, substituting (\ref{eq::value function})--(\ref{eq::the optimal gain}) into (\ref{eq::Bellman equation}) and multiplying both sides by positive definite matrix $P$ gives
\begin{equation}
	\label{eq::GARE}
	AP + PA^T + Q - P(C^TR^{-1}C - \gamma^{-2} L^T S L)P = 0
\end{equation}
which is exactly the game algebraic Riccati equation (GARE). Therefore, $K^*$ is the gain of $H_\infty$ filter, and the solving process mentioned in the previous section can be regarded as an extension of the traditional solution method of $H_\infty$ filter.

\section{Ternary Policy Iteration Algorithm for H-infinity Filter}
\label{sec::ternary policy iteration algorithm for H_infty filter}

\begin{algorithm}[tb]
	\caption{Ternary Policy Iteration for $H_\infty$ Filter}
	\label{TPI}
	\textbf{Input}: Initial parameters of neural networks\\
	\textbf{Parameter}: Learning rates of value function, filter gain and process noise are $\alpha_{\omega}$, $\alpha_{\theta}$ and $\alpha_{\eta}$\\
	\textbf{Output}: Gain $K(\theta)$ of filter
	\begin{algorithmic}[0] 
		\STATE Let $k = 0$.
		\STATE 0. Generate dataset $\mathcal{D}$ by applying $K(\theta^k)$ and $w(\tilde{x};\eta^k)$ to multiple agents
		\STATE 1. Given filter gain $K(\theta^k)$ and process noise $w(\tilde{x};\eta^k)$, update value function
		\begin{equation}
		\nonumber
		\omega^{k+1} = \omega^k - \alpha_{\omega}\frac{\partial L_{\omega}(\omega^k, \theta^k, \eta^k)}{\partial \omega^k}
		\end{equation}
		\STATE 2. Fix measurement noise $v(\tilde{x};\theta^k)$, update gain and process noise
		\begin{equation}
		\nonumber
		\theta^{k+1} = \theta^k - \alpha_{\theta}\frac{\partial L_{\theta}(\omega^{k+1}, \theta^k, \eta^k)}{\partial \theta^k}
		\end{equation}
		\begin{equation}
		\nonumber
		\eta^{k+1} = \eta^k - \alpha_{\eta}\frac{\partial L_{\eta}(\omega^{k+1}, \theta^k, \eta^k)}{\partial \eta^k}
		\end{equation}
		\STATE 3. Go back to 0 and $k \leftarrow k + 1$
	\end{algorithmic}
\end{algorithm}

In the previous section, we have derived a nonquadratic HJI equation corresponding to $H_\infty$ filter subject to bounded noise, which is a nonlinear partial differential equation. Its analytical solution is difficult to find. Inspired by our previous work \cite{c15} in optimal control, a ternary policy iteration (TPI) algorithm is proposed to solve the HJI equation resulting in reinforcement filter. Neural networks are employed to approximate value function, gain and noise to improve approximation abilities and avoid spending time in designing features like polynomial bases. An inner-loop iteration to evaluate the value function of given filter gain and noise is removed to reduce calculations. Besides, this method can deal with bounded noise and has the potential to handle non-affine or nonlinear systems.

The proposed TPI algorithm contains three updating phases and three relevant loss functions with three approximate networks named value network $V(\tilde{x};\omega)$, gain network $K(\theta)$ and noise network $w(\tilde{x};\eta)$, whose parameters are denoted as $\omega$, $\theta$ and $\eta$. Substituting networks and the fixed measurement noise
\begin{equation}
	\nonumber
	v(\tilde{x};\theta) = -\bar{V}\tanh\left(\frac{1}{2\gamma^2}(R\bar{V})^{-1} K^T(\theta)\frac{\partial V(\tilde{x})}{\partial \tilde{x}}\right)
\end{equation} 
to Hamiltonian (\ref{eq::Hamiltonian with bounded noise}) or (\ref{eq::Bellman equation}) gives the approximate Hamiltonian
\begin{equation}
	\nonumber
	\begin{array}{c}
	\displaystyle{H(\tilde{x}, \theta, \eta, \omega) = l(\tilde{x}, K(\theta), v(\tilde{x};\theta), w(\tilde{x};\eta))} \\
	\displaystyle{+ \frac{\partial V(\tilde{x};\omega)}{\partial \tilde{x}^T}((A - K(\theta)C)\tilde{x} + w(\tilde{x};\eta) - K(\theta)v(\tilde{x};\theta))}
	\end{array}
\end{equation}
which is utilized to construct three loss functions to update parameters of approximate networks. 

\begin{figure}[t] 
	\centering
	\includegraphics{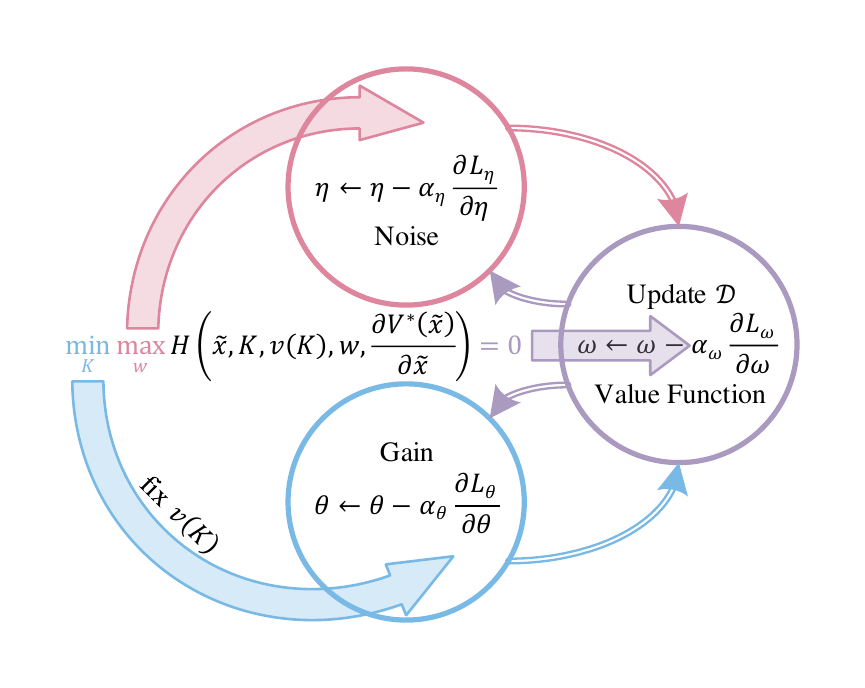}
	\caption{Ternary Policy Iteration Algorithm}
	\label{fig::Ternary Policy Iteration Algorithm}
\end{figure}

Different from other policy iteration algorithms, TPI only updates parameters once in each phase of an iteration by gradient descent (GD) method to decrease designed loss functions. The pseudo-code of the TPI algorithm is presented in \textbf{Algorithm \ref{TPI}}, and corresponding iteration procedure is shown in Fig.~\ref{fig::Ternary Policy Iteration Algorithm}. The specific operations of each phase are as follows.

1) Value Function Update Phase

In policy evaluation step of traditional policy iteration algorithm, value function is updated to evaluate given filter gain and noise by solving the Bellman equation (\ref{eq::Hamiltonian with bounded noise}) or (\ref{eq::Bellman equation}) for all states in entire state set $\Omega$. In order to reduce the amount of calculation in our value function update phase, first define the value loss function on the subset $\mathcal{D}\subset\Omega$ we care about
\begin{equation}
	\nonumber
	L_{\omega}\left(\omega^k, \theta^k, \eta^k\right) = \mathbb{E}_{\tilde{x}\in\mathcal{D}} \left[\left|H(\tilde{x}, \theta^k, \eta^k, \omega^k)\right|\right]
\end{equation}
where $\mathcal{D}$ is generated and refreshed by multiple agents.

The parameters of value function is revised by GD method to make the expectation of Hamiltonian on the subset $\mathcal{D}$ gradually tend to zero instead of letting the Bellman equation (\ref{eq::Hamiltonian with bounded noise}) or (\ref{eq::Bellman equation}) always hold in $\Omega$.
\begin{equation}
	\label{eq::value function update phase}
	\omega^{k+1} = \omega^k - \alpha_{\omega}\frac{\partial L_{\omega}(\omega^k, \theta^k, \eta^k)}{\partial \omega^k}
\end{equation}

Although the measurement noise $v$ has a term of value function, $v$ is regarded as a constant in value loss function whose gradient about $\omega^k$ is not calculated.

2) Gain Update Phase

In existing dynamic programming methods, policy improvement is completed by explicitly minimizing Hamiltonian after policy evaluation. But this may not suitable for general non-affine systems and complex utilities like the nonquadratic one (\ref{eq::nonquadratic utility}) considering bounded noise. In order to simplify the calculation of the minimum value, define the gain loss function analogous to the value loss function as
\begin{equation}
	\nonumber
	L_{\theta}\left(\omega^{k+1}, \theta^k, \eta^k\right) = \mathbb{E}_{\tilde{x}\in\mathcal{D}} \left[H(\tilde{x}, \theta^k, \eta^k, \omega^{k+1})\right]
\end{equation}

Then, the parameters of filter gain is revised by decreasing the expectation of Hamiltonian to cope with non-affine property and complex utility function
\begin{equation}
	\label{eq::gain update phase}
	\theta^{k+1} = \theta^k - \alpha_{\theta}\frac{\partial L_{\theta}(\omega^{k+1}, \theta^k, \eta^k)}{\partial \theta^k}
\end{equation}

Because the parameters of filter gain is included in the expression of the fixed measurement noise $v(\tilde{x};\theta^k)$, its gradient about $\theta^k$ should be reflected in the gradient descent process.

3) Noise Update Phase

Contrary to the optimization direction of gain update phase, the objective of noise update phase is to gradually increase the expectation of Hamiltonian, i.e.
\begin{equation}
	\label{eq::noise update phase}
	\eta^{k+1} = \eta^k - \alpha_{\eta}\frac{\partial L_{\eta}(\omega^{k+1}, \theta^k, \eta^k)}{\partial \eta^k}
\end{equation}
where the noise loss function is defined as
\begin{equation}
	\nonumber
	L_{\eta}\left(\omega^{k+1}, \theta^k, \eta^k\right) = \mathbb{E}_{\tilde{x}\in\mathcal{D}} \left[-H(\tilde{x}, \theta^k, \eta^k, \omega^{k+1})\right]
\end{equation}
which is the opposite number of gain loss function resulting in an synchronous updating algorithm.

\section{Illustrative Examples}
\label{sec::illustrative examples}

In this section, the proposed TPI method is first applied to a linear system to show that it converges to the optimal solution to $H_\infty$ filter. Then, it is tested on a setting of bounded noise to verify its effectiveness. The dynamics employed is the following bicycle model with two degrees of freedom:
\begin{equation}
	\label{eq::bicycle model}
	\setlength{\arraycolsep}{0.5mm}
	\left\{
	\begin{array}{c}
		\dot{x} = \left[\begin{array}{cc}
			\frac{k_f + k_r}{m u} & \frac{a k_f - b k_{r}}{m u^2} - 1 \\
			\frac{a k_f - b k_r}{I_{zz}} & \frac{a^2 k_f + b^2 k_r}{u I_{zz}}
		\end{array}\right] x + \left[\begin{array}{c}
			-\frac{k_f}{m u} \\
			-\frac{a k_f}{I_{zz}}
		\end{array}\right] \delta + w \\
		y = \left[\begin{array}{cc}
			\frac{k_f + k_r}{m} & \frac{a k_f - b k_r}{m u} \\
			0 & 1
		\end{array}\right] x + \left[\begin{array}{c}
			-\frac{k_f}{m} \\
			0
		\end{array}\right] \delta + v \\
		z = \left[\setlength{\arraycolsep}{1mm}\begin{array}{cc}
			1 & 0 \\
			0 & 1
		\end{array}\right] x
	\end{array}\right.
\end{equation}
where $x = \left[\beta \quad \omega_r\right]^T$ is the state of the system, $y = \left[a_y \quad \omega_r\right]^T$ is the measurement output, $z$ is the objective output, $\beta$ is the sideslip angle of the center of gravity (CG), $\omega_r$ is the yaw rate, $a_y$ is the lateral acceleration, the control variable $\delta$ is steering angle which is not related to noise and can be obtained directly, and $w$ and $v$ are process and measurement noise. Other parameters can be found in \cite{c15} except that the longitudinal velocity $u$ is set as 20 m/s.

\begin{figure}[t] 
	\centering
	\makebox{\parbox{3in}{\includegraphics[scale=0.5]{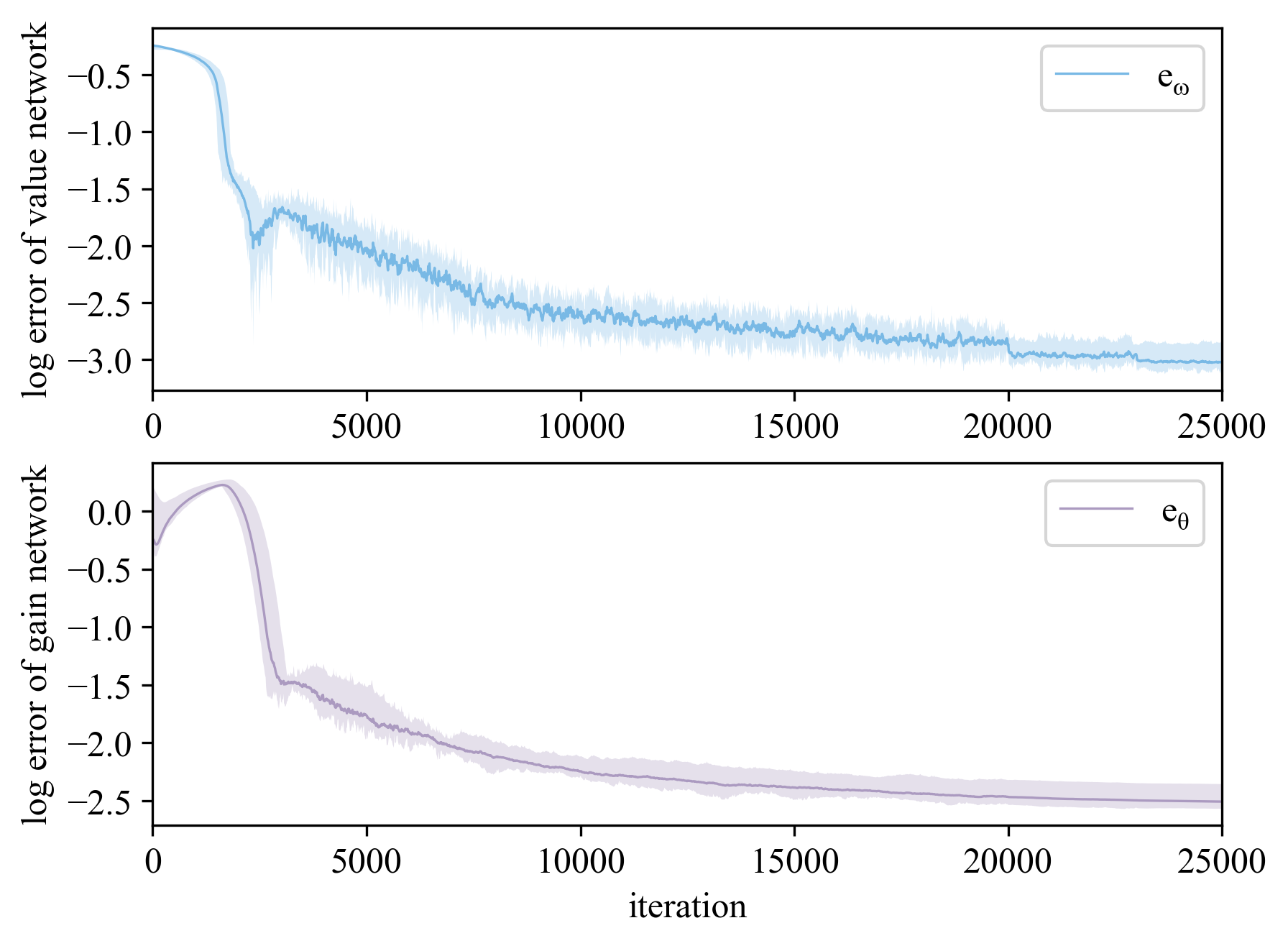}}}
	\caption{Relative errors of the weights of networks}
	\label{fig::relative errors of the weights of networks}
\end{figure}

\subsection{Not Involving the Bounded Property of Noise}

Set value network as $V(\tilde{x}; \omega) = \tilde{x}^T P^{-1} \tilde{x} = \omega^T \sigma(\tilde{x})$, gain network as $K(\theta) = \theta$, noise networks as $w(\tilde{x}; \eta) = \eta^T \tilde{x}$ and $v(\tilde{x}; \theta) = -\frac{1}{2\gamma^2} R K^T(\theta) \frac{\partial V(\tilde{x})}{\partial \tilde{x}}$, where the feature function and weights of networks are
\begin{equation}
	\label{eq::feature and weights of network}
	\begin{array}{c}
	\sigma(\tilde{x}) = \left[\tilde{x}_1^2 \quad \tilde{x}_1 \tilde{x}_2 \quad \tilde{x}_2^2\right]^T \\
	\omega \in \mathbb{R}^{3 \times 1}, \theta \in \mathbb{R}^{2 \times 2}, \eta \in \mathbb{R}^{2 \times 2}
	\end{array}
\end{equation}

In the utility function $l(\tilde{x}, K, v, w)$, set $Q = 20I$, $R = 10I$, $S = I$ and $\gamma = 1$. Then, the corresponding GARE (\ref{eq::GARE}) can be solved directly, and the real values of the weights of value network and gain network can be obtained:
\begin{equation}
	\nonumber
	\begin{array}{c}
		\omega^* = \left[8.8898 \quad -0.1247 \quad 0.5225\right]^T \\
		\theta^* = \left[\begin{array}{cc}
			-1.3646 & 0.0013 \\
			0.0367 & 0.1916
		\end{array}\right]
	\end{array}
\end{equation}

Define the relative errors of the weights of value network and gain network as
\begin{equation}
	\label{eq::relative error of weights}
	\begin{array}{c}
		\displaystyle{e_{\omega} = \frac{\left\|\omega-\omega^*\right\|_2}{\left\|\omega^*\right\|_2}} \\
		\displaystyle{e_{\theta} = \frac{\left\|\theta-\theta^*\right\|_F}{\left\|\theta^*\right\|_F}}
	\end{array}
\end{equation}

Set the number of agents as 64, the initial learning rate of all networks is 0.05. Gradually reducing learning rates during learning process can greatly reduce the relative errors. As shown in Fig.~\ref{fig::relative errors of the weights of networks} averaged over 10 runs, the relative errors of weights are reduced to around $10^{-3}$ after 25000 iterations. As a result, the TPI algorithm converges to the optimal solution to $H_\infty$ filter for the linear system.

\subsection{Considering the Bounded Property of Noise}

Utilize the linear plant (\ref{eq::bicycle model}) with bounded noise, where $|w_1| \leq 0.01$, $|w_2| \leq 0.05$, $|v_1| \leq 0.01$ and $|v_2| \leq 0.05$. In the utility function of reinforcement filter, set $Q = 0.2I$, $R = 0.1I$, $S = I$ and $\gamma = 1$, which ensures that the magnitude of noise terms is the same as that in the previous example. Gain network is also set as $K(\theta) = \theta$, and 3-layer fully connected networks are employed to approximate value function and process noise, where both of networks consist of 64 neurons per layer, the activation function of the first two layers is SELU, and the activation function of the last layer is Tanh. The learning rate of three networks is $10^{-2}$, and Adam method is implemented to update parameters of networks.

Use the estimated and true values of the state to define the root mean square (RMS) of state estimation error as follows
\begin{equation}
	\label{eq::RMS}
	\begin{array}{c}
	{\rm RMS}_{\beta} = 10^4 \displaystyle{\sqrt{\int_0^T (\hat{\beta}(t) - \beta(t))^2 dt}} \\
	{\rm RMS}_{\omega_r} = 10^4 \displaystyle{\sqrt{\int_0^T \left(\hat{\omega}_r(t) - \omega_r(t)\right)^2 dt}}
	\end{array}
\end{equation}

Set steering angle as $\delta(t) = \frac{0.5\pi}{180} \sin (\frac{2}{3}\pi t)$. Apply different filters to repeat filtering test 100 times under different bounded noise distributions $w_i = 2\bar{w}_i X - \bar{w}_i$ and $v_i = 2\bar{v}_i X - \bar{v}_i$, where $X \in [0, 1]$ is a random variable, Kalman filter employs the true covariance matrix of noise, the duration of each test is 25 s and sampling frequency is 200 Hz. Corresponding average value of RMS of state estimation error is summarized in Table \ref{tab::comparison of filtering effect with bounded noise}. 

\begin{table}[t]
	\caption{Comparison of Filtering Effect with Bounded Noise}
	\label{tab::comparison of filtering effect with bounded noise}
	\begin{center}
		\setlength{\tabcolsep}{1.0mm}{
			\begin{tabular}{c c c c c c c}
				\toprule
				Noise & \multicolumn{2}{c}{Reinforcement Filter} & \multicolumn{2}{c}{$H_{\infty}$ Filter} & \multicolumn{2}{c}{Kalman Filter}\\
				\midrule
				$X$ & ${\rm RMS}_{\beta}$ & ${\rm RMS}_{\omega_r}$ & ${\rm RMS}_{\beta}$ & ${\rm RMS}_{\omega_r}$ & ${\rm RMS}_{\beta}$ & ${\rm RMS}_{\omega_r}$ \\
				\midrule
				U(0, 1) & 2.157 & 28.99 & 2.472 & 29.30 & 2.166 & 27.52 \\
				Beta(2, 2) & 1.675 & 22.23 & 1.909 & 22.65 & 1.675 & 21.48 \\
				Triang(0,1,0.6) & 1.609 & 30.79 & 1.834 & 30.30 & 1.632 & 33.40 \\
				Beta(4, 2) & 2.662 & 116.5 & 3.008 & 112.6 & 3.034 & 136.5 \\
				\bottomrule
		\end{tabular}}
	\end{center}
\end{table}

It can be found that for the first two distributions, the filtering effect of Kalman filter is better than the other two filters because Kalman filter is the best linear estimation for zero-mean noise \cite{c1}. However, for the noise with non-zero mean, the effect of reinforcement filter solved by the TPI algorithm is superior to Kalman filter, and its effect is slightly better than that of $H_\infty$ filter. Therefore, the effectiveness of the proposed filter against bounded noise has been validated.

\section{Conclusions}
\label{sec::conclusions}

Reinforcement solutions to $H_\infty$ filter copping with bounded noise are presented in this study. A ternary policy iteration algorithm with three updating phases and three networks is proposed to solve the nonquadratic HJI equation matched with reinforcement filter. Measurement noise is fixed in the updating process to enlarge the convexity with respect to filter gain, which handles the non-affine property of dynamics and nonquadratic utility depicting bounded noise. The first illustrative example verifies the convergence of the proposed algorithm compared with the gain matrix of $H_\infty$ filter. The second example demonstrates the effectiveness and accuracy of the reinforcement filter when considering bounded noise.


\end{document}